\documentclass[aps,pra,twocolumn,showpacs]{revtex4-1}
\usepackage{hyperref}
\usepackage{amsmath,amssymb}
\usepackage{graphicx}
\usepackage{mathtools}
\usepackage{paralist}

\usepackage{ntheorem}
\newtheorem{thm}{Theorem}

\newcommand{\ket}[1]{| #1 \rangle}
\newcommand{\bra}[1]{\langle #1 |}
\newcommand{\dyad}[2]{ | #1 \rangle  \langle #2 | }
\newcommand{\av}[1]{ \langle #1 \rangle}

\newcommand{\mean}{\mathrm{E}}
\newcommand{\var}{\mathrm{Var}}
\newcommand{\abs}[1]{ | #1 |}
\newcommand{\Imm}{\mathrm{Im}}

\newcommand{\eqqref}[1]{Eq.~\eqref{#1}}
\newcommand{\figref}[1]{Fig.~\ref{#1}}
\newcommand{\secref}[1]{Sec.~\ref{#1}}
\newcommand{\thmref}[1]{Thm.~\ref{#1}}

\newcommand{\erf}{\mathrm{erf}}
\newcommand{\erfi}{\mathrm{erfi}}

\begin{document}
\title{Stochastic excitation during the decay of a two-level emitter subject to homodyne and heterodyne detection}
\author{Anders Bolund} \author{Klaus M\o lmer}
\affiliation{Lundbeck Foundation Theoretical Center for Quantum System
  Research, Department of Physics and Astronomy, University of Aarhus,
  DK-8000 Aarhus C, Denmark}

\date{\today}

\begin{abstract}
  We study the dynamics of an atomic two-level system decaying by
  spontaneous emission of light. Subject to continuous detection of
  the radiated field, the system tends with certainty to the ground
  state in the long time limit, but at initial times the excited state
  population exhibits non-trivial stochastic behavior. Employing
  methods from Ito calculus, we characterize this behavior, and we
  show, for example, that the emitter, as a result of in-phase
  homodyne measurements, may become fully excited during the decay
  process while heterodyne and out of phase homodyne measurements do
  not drive the atom completely into the excited state.
\end{abstract}

\maketitle

\section{Introduction}

With the ability to control single emitters in the form of trapped
ions, atoms, NV-centers, Josephson qubits, etc., the real time
dynamics of single quantum systems subject to measurement constitutes
a very active research field with wide perspectives for fundamental
investigations as well as for quantum metrology, quantum computing,
and quantum feedback and control.

In quantum optics light emitting systems are often studied by
subjecting the radiation to photon counting, heterodyne or homodyne
measurements. In case of continuous measurements, the conditioned
quantum state of the emitter evolves according to a stochastic master
equation (SME) or, if the state remains pure as a result of the
measurements, a stochastic Schr\"{o}dinger equation (SSE)
\cite{wiseman2009,gardiner2000,dalibard1992,carmichael1994}. The SME
and SEE explicitly reveal the fluctuation and noise characteristics of
the emission process and the detection back-action, while averaging
over an ensemble of measurement histories leads to the conventional
Lindblad master equation describing the state of an unobserved emitter
subject to decoherence and decay. Dynamical equations similar to the
measurement SSEs also arise in the description and efficient
simulation of open systems \cite{dalibard1992, breuer2002} as well as
in collapse models aiming to shed light on the measurement problem
\cite{ghirardi1986, gisin1992}.

\begin{figure}[htbp]
\centering
\includegraphics[scale = 1]{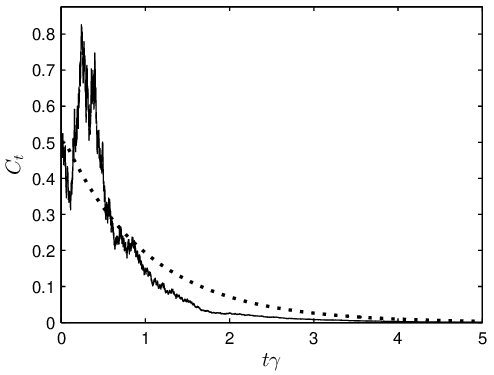}
\caption{\small A single realization of the upper level population
  $C_{t}$ (solid curve), exhibiting transient fluctuations that exceed
  the initial value $C_{0} = 0.5$ during homodyne detection. The
  dotted curve shows the exponentially decaying, unconditioned excited
  state probability $\rho_{ee}(t)$ with $\rho_{ee}(0) = C_{0}$.}
\label{C_t_pex}
\end{figure}

In this paper we study the state of a decaying two-level quantum
system subject to continuous, unit efficiency heterodyne or homodyne
detection of the emitted light. For photon counting it is well known
that the state of a two-state emitter experiences continuous decay in
periods with no photodetection events, while a photon count event is
accompanied by a quantum jump of the system into the ground state
\cite{dalibard1992}. For heterodyne and homodyne detection, on the
other hand, the continuous measurements of the intensity interference
signal between the source field and a local oscillator field cause the
excitation of the emitter to undergo random, diffusive time evolution
\cite{wiseman1993,hofmann1998}. Such behavior is illustrated in
\figref{C_t_pex}, where the excited state population (solid curve) of
a two-level emitter under homodyne detection is compared with the
exponential decay (dotted curve) of the unobserved system. In this
article, we use techniques from Ito calculus to analytically
characterize different aspects of the measurement-induced dynamics of
the light emitter. In particular, we determine the probability that
the emitter during its decay attains any specified degree of
excitation, and we calculate the mean time the observed atom spends
above its initial excitation as well as the average time needed for
the system to reach different excited state populations below the
initial value.

In \secref{tls}, we introduce the Lindblad master equation and the
stochastic wave function descriptions associated with photon counting
and heterodyne and homodyne detection. For heterodyne and homodyne
detection, we determine the resulting stochastic differential
equations for the excited state population of the emitter. In
\secref{general_hom_detection}, we present numerical simulations of
homodyne detection with arbitrary local oscillator phases. Focusing in
Secs.~\ref{excitation_dynamics} and \ref{char_times} on heterodyne
detection and homodyne detection with the local oscillator initially
in phase with the emitter dipole, we derive adjoint Fokker-Planck-type
equations for interesting probabilistic and average properties of the
conditioned atom. In \secref{excitation_dynamics}, we thus determine
the probability that the excited state population reaches the upper
endpoint of an interval before the lower one, and we heuristically
discuss the attainability of zero and unit excitation. This permits
analytical evaluation of the probability for the excited state
population to reach any level. In \secref{char_times}, we obtain an
expression for the average time it takes the upper level population to
leave definite intervals, and we use this result to rigorously discuss
the attainability of zero and unit excited state occupation. We also
determine the mean time the emitter subject to detection spends above
the initial excited state population as well as the average time it
takes the upper level population to attain a specified value below the
initial excitation. \secref{conclusion} concludes the paper.

\section{Stochastic description of a two-level quantum emitter}
\label{tls}

We consider a two-level quantum system with excited state $\ket{e}$
decaying into the ground state $\ket{g}$ by spontaneous emission of
electromagnetic radiation. The time evolution is treated in the
electric dipole, rotating wave, and Born-Markov approximations, and if
no measurements are performed on the emitted field, the system will,
as a result of the interaction-induced entanglement with the quantized
radiation field, evolve into a mixed state obeying the interaction
picture master equation (we set $\hbar = 1$ throughout)
\cite{wiseman2009,breuer2002}
\begin{align}
d \rho(t) = \gamma \bigl( \sigma_{-} \rho(t) \sigma_{+} - \frac{1}{2}\{\sigma_{+} \sigma_{-},\rho(t) \} \bigr) dt,
\label{ME}
\end{align}
where $\gamma$ is the decay rate, $\sigma_{-} = \dyad{g}{e}$ is the
lowering operator, and $\sigma_{+} = \dyad{e}{g}= \sigma_{-}^\dagger$
is the raising operator of the two-level emitter.

The solution to the master equation \eqqref{ME} yields the well known
exponential decay of the excited level population and the coherences,
\begin{align}
& \rho_{ee}(t) = 1 - \rho_{gg}(t) = \rho_{ee}(0) e^{-\gamma t } \label{rho_ee} \\
& \rho_{eg}(t) = \rho_{ge}^{*}(t) = \rho_{eg}(0) e^{- \frac{1}{2} \gamma t }. \nonumber
\end{align}

When the light emitted from the decaying system is subject to
continuous photon counting, homodyne or heterodyne detection, the
atom-field entanglement mediates a detection back-action on the system
state which, in the case of unit efficiency measurements, restores the
pure state character of the emitter, provided it is initially prepared
in a pure state.

For continuous photon counting \cite{dalibard1992,carmichael1994}, the
evolution is split up into no-jump and jump dynamics, such that the
pure state $\ket{\psi(t)} = c_{e}(t) \ket{e} + c_{g}(t) \ket{g}$ of
the emitter during intervals with no photon detection experiences a
damping rate of $\gamma/2$, and the upper level population $C_{t}
\equiv \abs{c_{e}(t)}^{2}$ evolves as $C_{t} = C_{0} e^{-\gamma t }/(
1- C_{0} + C_{0} e^{-\gamma t })$. Associated with the detection of a
photon, this non-exponential decay of the excitation is with
probability $\gamma C_{t} dt$ interrupted at time $t$ by a jump of the
state vector into the ground state, where the system remains in the
absence of an external driving field. The average excited state
population over many realizations of this dynamics is given by
\eqqref{rho_ee}: jumps occur with the rate $\gamma \rho_{ee}(t)$, and
on average, a fraction $\rho_{ee}(0)$ of the emitters will reach the
ground state by a jump, while the remaining fraction $\rho_{gg}(0)$
enter the ground state by continuous, exponential no-jump
evolution. For both the jump and no-jump dynamics, the excited state
population of the two-level emitter is always lower than in the
initial state.

For homodyne detection with local oscillator phase $\phi$, the system
evolves according to the interaction picture stochastic
Schr\"{o}dinger equation (SSE)
\cite{wiseman2009,gardiner2000,wiseman1993}
\begin{align}
d\ket{\psi(t)} = \frac{1}{2} \gamma \Bigl( -\sigma_{+}\sigma_{-} & + \av{\sigma_{+}e^{i\phi} + \sigma_{-}e^{-i\phi}}\sigma_{-}e^{-i\phi} \nonumber \\
& - \frac{1}{4} \av{\sigma_{+}e^{i\phi} + \sigma_{-}e^{-i\phi}}^{2} \Bigr) \ket{\psi(t)} dt \nonumber \\
+ \sqrt{\gamma} \Bigl( \sigma_{-}e^{-i\phi} & - \frac{1}{2}\av{\sigma_{+}e^{i\phi} + \sigma_{-}e^{-i\phi} } \Bigr) \ket{\psi(t)} dW_{t},
\label{SSE_hom}
\end{align}
where $\av{\ldots}$ denotes the quantum mechanical expectation value
$\bra{\psi(t)} \ldots \ket{\psi(t)}$. The independent, Gaussian
distributed Wiener increments $\{ dW_{t} \}_{t \geq 0}$ with mean
$\mean(dW_{t}) = 0$ and variance $\var(dW_{t}) = dt$ represent the
random outcomes of the homodyne detection through the appropriately
scaled differential measurement current $dq_{t} = \gamma \av{
  \sigma_{+}e^{i\phi} + \sigma_{-}e^{-i\phi} } dt + \sqrt{\gamma}
dW_{t}$ and obey the Ito calculus rules $dW_{t}^{2} = 0$ and $dW_{t}
dt = 0$ \cite{gardiner2000}.

The corresponding heterodyne SSE reads \cite{wiseman2009,wiseman1993}
\begin{align}
d\ket{\psi(t)} = \gamma \Bigl( -\frac{1}{2} \sigma_{+}\sigma_{-} & + \av{\sigma_{+}}\sigma_{-} - \frac{1}{2} \av{\sigma_{+}}\av{\sigma_{-}} \Bigr) \ket{\psi(t)} dt \nonumber \\
& + \sqrt{\gamma} \Bigl( \sigma_{-} - \av{\sigma_{-}} \Bigr) \ket{\psi(t)} dZ_{t},
\label{SSE_het}
\end{align}
where the complex Wiener noise $dZ_{t} = \frac{1}{\sqrt{2}} (dW_{t,x}
+ i dW_{t,y})$ is defined in terms of two real-valued, independent
Wiener increments $dW_{t,x}$ and $dW_{t,y}$ with $dW_{t,x}dW_{t,y} =
0$. $dZ_{t}$ thus obeys $dZ_{t}^{2} = 0$ and $dZ_{t}^{*}dZ_{t} =
dt$. The complex differential measurement current is $dq_{t} = \gamma
\av{\sigma_{-}}dt + \sqrt{\gamma} dZ_{t}$.

Upon calculating $d(\dyad{\psi(t)}{\psi(t)} ) = d\ket{\psi(t)}
\bra{\psi(t)} + \ket{\psi(t)} d\bra{\psi(t)} + d\ket{\psi(t)}
d\bra{\psi(t)}$ from \eqqref{SSE_hom} or \eqqref{SSE_het} and using
the Ito calculus rules, we verify that the ensemble average $\mean(
\dyad{\psi(t)}{\psi(t)} )$ fulfills the master equation \eqqref{ME}.

Applying the expansion $\ket{\psi(t)} = c_{e}(t)\ket{e} +
c_{g}(t)\ket{g}$ in \eqqref{SSE_hom} and \eqqref{SSE_het}, we obtain
coupled equations for $c_{e}(t)$ and $c_{g}(t)$. Using the Ito
calculus rules, and the normalization $\abs{c_{e}(t)}^{2} +
\abs{c_{g}(t)}^{2} = 1$, we find a stochastic equation for the excited
state population $C_{t} \equiv \abs{c_{e}(t)}^{2}$. Similarly, by
taking the differential of $\varphi_{t} = \Imm \bigl[ \ln\bigl(
c_{g}^{*}(t)c_{e}(t) \bigr) \bigr]$, we find an equation for the phase
difference $\varphi_{t} = \varphi_{e}(t) - \varphi_{g}(t)$ between
$c_{e}(t) = \abs{c_{e}(t)}e^{i\varphi_{e}(t)}$ and $c_{g}(t) =
\abs{c_{g}(t)}e^{i\varphi_{g}(t)}$.

For heterodyne detection, the resulting equations read
\begin{align}
dC_{t} & = - \gamma C_{t}dt - \frac{2}{\sqrt{2}} \sqrt{\gamma} \sqrt{C_{t}^{3}(1-C_{t})} dW_{t} \nonumber \\
& \equiv A(C_{t})dt + B(C_{t})dW_{t} \label{SSE_C_het}
\end{align}
\begin{align*}
d\varphi_{t} & = - \sqrt{\frac{\gamma}{2}} \sqrt{\frac{C_{t}}{1-C_{t}}} \bigl( \sin(\varphi_{t}) dW_{t,x} + \cos(\varphi_{t}) dW_{t,y} \bigr),
\end{align*}
where we have defined $dW_{t} = \cos(\varphi_{t}) dW_{t,x} -
\sin(\varphi_{t}) dW_{t,y}$, which is seen to be a standard Wiener
increment upon using the properties of $dW_{x,t}$ and $dW_{y,t}$ and
their independence of $\varphi_{t}$. Consequently, \eqqref{SSE_C_het}
constitutes a closed equation on standard Ito form for $C_{t}$.

For homodyne detection we get instead the two coupled equations
\begin{align}
dC_{t} & = - \gamma C_{t}dt - 2 \sqrt{\gamma} \cos( \tilde{\varphi}_{t} ) \sqrt{C_{t}^{3}(1-C_{t})} dW_{t}
\label{SSE_C_hom} \\
d\tilde{\varphi}_{t} & = \gamma \frac{2C_{t}^{2}-C_{t}}{2(1-C_{t})} \sin( 2\tilde{\varphi}_{t} ) dt
- \sqrt{\gamma} \sin( \tilde{\varphi}_{t} ) \sqrt{\frac{C_{t}}{1-C_{t}}} dW_{t},
\label{SSE_phi_hom}
\end{align}
where $\tilde{\varphi}_{t} = \varphi_{t} - \phi$ and where we observe
from \eqqref{SSE_C_hom} that the initial phase $\tilde{\varphi}_{0}$
may be restricted to $[0,\frac{\pi}{2}]$ (by absorbing a possible
minus sign in the definition of $dW_{t}$).

For arbitrary local oscillator phases, the time evolution of $C_{t}$
under homodyne detection is generally coupled to that of the phase
variable $\tilde{\varphi}_{t}$. However, if initially the emitter
dipole is in phase with the local oscillator and $\tilde{\varphi}_{0}
= 0$, \eqqref{SSE_phi_hom} shows that $\tilde{\varphi}_{t} = 0$ and
thus $\cos( \tilde{\varphi}_{t} ) = 1$ for all future times $t > 0$.
Since the magnitude of the measurement back-action induced by the
diffusion term in \eqqref{SSE_C_hom} is largest when
$\abs{\cos(\tilde{\varphi}_{t})} = 1$, we will refer to this case as
\emph{optimal} homodyne detection and we get a closed equation for
$C_{t}$
\begin{align}
dC_{t} & = - \gamma C_{t}dt - 2 \sqrt{\gamma} \sqrt{C_{t}^{3}(1-C_{t})} dW_{t} \nonumber \\
& \equiv A(C_{t})dt + B(C_{t}) dW_{t}.
\label{SSE_C_hom_opt}
\end{align}

\figref{C_t_pex} shows a single realization of \eqqref{SSE_C_hom_opt}
found by numerical simulation using the Milstein scheme
\cite{kloeden1995}. Unlike the exponentially decaying excitation of
the unobserved atom, the excited state population conditioned on the
optimal homodyne detection signal reaches a maximum value well above
the initial excitation. It is the purpose of this paper to
characterize the stochastic dynamics of the excited state
population. In particular, we shall determine the probability
$P_{T_{u}<\infty}(y)$ that the excitation $C_{t}$ of the emitter (in
finite time $T_{u} < \infty$) reaches different population levels $u$
above the initial value $C_{0} = y$ for the different detection
schemes.

\section{Numerical simulation of general homodyne detection}
\label{general_hom_detection}

For general homodyne detection, with a local oscillator phase which
may differ from the initial phase of the dipole, the coupled equations
\eqref{SSE_C_hom} and \eqref{SSE_phi_hom} for the population and phase
variables $C_t$ and $\tilde{\varphi}_t$ cannot be approached
analytically and in this section we estimate $P_{T_{u}<\infty}(y)$ by
numerical simulations of \eqqref{SSE_hom}.

\begin{figure}[htbp]
\centering
\includegraphics[scale = 1]{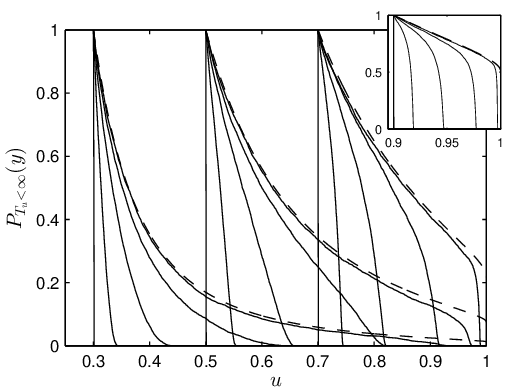}
\caption{\small The probabilities $P_{T_{u} < \infty}(y)$ that the
  upper level population $C_{t}$ attains the value $u$ above its
  initial value $C_{0} =y = 0.3,0.5,0.7$ are shown as groups of curves
  from left to right in larger figure while the results for $C_{0} = y
  = 0.9$ are shown in the smaller figure. Each group of curves
  presents calculations carried out with initial homodyne detection
  phases $\tilde{\varphi}_{0}/\frac{\pi}{2} = 1,0.7,0.5,0.3,0.1$ (from
  left to right), converging towards the analytical result for optimal
  homodyne detection with $\tilde{\varphi}_{0} = 0$ (dashed curve).}
\label{P_Tuy_num}
\end{figure}

In \figref{P_Tuy_num}, we present data for simulations, where we begin
with excited state population $C_{0} = y = 0.3, 0.5, 0.7,0.9$, and
where the homodyne local oscillator phase initially differs from the
dipole phase by $\tilde{\varphi}_{0}/\frac{\pi}{2} = 0.1, 0.3, 0.5,
0.7,1$. For each initial population and phase, we propagate $5000$
independent solutions of \eqqref{SSE_hom} and register the fraction of
trajectories $C_{t}$ that exceed different values $u$ above the
initial value $C_{0} = y$. This fraction is thus a numerical estimate
for the desired probability $P_{T_{u}<\infty}(y)$ that the emitter
starting with an excited state population $y$ reaches the excitation
level $u \geq y$ in finite time.

For the simulations, we use an explicit 1.0 strong order Ito-Taylor
scheme \cite{kloeden1995} with a global pathwise error of
$\mathcal{O}(\Delta t)$, where $\Delta t$ is the temporal step
size. All simulations have been carried out on the time interval $[0
\gamma^{-1},5 \gamma^{-1}]$ and with a step size of $\Delta t =
10^{-4} \gamma^{-1}$. In the case of optimal homodyne detection with
$\tilde{\varphi}_{0} = 0$, we have for all values of $u \in [0,1[$
verified that our simulations of $P_{T_{u}<\infty}(y)$ accurately to
within $\pm 0.02$ reproduce our analytical derivation,
\eqqref{P_T_u_<_inf}.

Each simulated emitter achieves with certainty the initial excited
state population and, depending on the homodyne detection phase, it
reaches with different probabilities excitation levels above the
initial value. The results of our simulations are reported in
\figref{P_Tuy_num} as sets of curves decreasing from unit probability
beyond different upper state population levels $u$. The upper dashed
curves are exact results for in-phase homodyne detection. We see from
\eqqref{SSE_phi_hom} that an initially vanishing phase difference
$\tilde{\phi}_{t}$ remains zero for all times, and that this choice of
phase leads to the strongest stochastic back-action in
\eqqref{SSE_C_hom}. In \figref{P_Tuy_num}, this is reflected by higher
excitation levels during the atomic decay process. For
$\tilde{\varphi}_{0} = \frac{\pi}{2}$, we see from \figref{P_Tuy_num}
that $C_{t}$ never exceeds its initial value during the homodyne
probing of the emitted radiation. For smaller values of the detection
phase, $\tilde{\varphi_{0}}/\frac{\pi}{2} = 0.1,0.3,0.5,0.7$, the atom
may be become excited, but the level of excitation decreases with
increasing phase, and none of our simulations lead to a value of
$C_{t}$ exceeding $0.999$.

\section{Probabilistic description of excitation dynamics under heterodyne and optimal homodyne detection}
\label{excitation_dynamics}

In case of heterodyne and optimal homodyne detection, the excited
state population $C_{t}$ evolves independently of the phase variable
according to the closed equations \eqref{SSE_C_het} and
\eqref{SSE_C_hom_opt}, respectively. These population equations,
however, cannot be reduced to linear equations and are thus not
explicitly solvable by the reduction method \cite{kloeden1995}.

Fortunately, Eqs.~\eqref{SSE_C_het} and \eqref{SSE_C_hom_opt} may be
analytically investigated using standard results from Ito calculus. To
begin the analysis, we observe that once $C_{t}$ hits $0$, the
excitation is irretrievably lost since $dC_{t}|_{C_{t} = 0} = 0$ and
the system remains in the ground state. If, on the other hand, $C_{t}$
reaches unity, $dC_{t}|_{C_{t} = 1} = -\gamma dt$, and the atom with
certainty decays away from the excited state again. With these
boundary conditions a standard analysis \cite{ikeda1989,gihman1972}
implies that Eqs.~\eqref{SSE_C_het} and \eqref{SSE_C_hom_opt} have
pathwise unique, (almost surely) continuous, and Markovian solutions
on $[0,1]$, as also required by the conservation of probability and
the Markov approximation used in the derivation of the SSEs.

\begin{figure}[htbp]
\centering
\includegraphics[scale = 1]{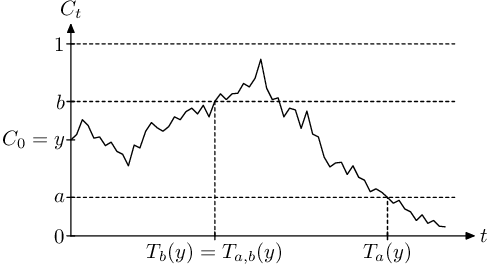}
\caption{\small First hitting times $T_{a}(y)$ and $T_{b}(y)$ of the
  levels $a$ and $b$, respectively, and the first exit time
  $T_{a,b}(y)$ of the interval $]a,b[$ for a particular realization of
  the upper level population $C_{t}$ given in \eqqref{SSE_C_het} or
  \eqqref{SSE_C_hom_opt} with $C_{0} = y \in [a,b]$. Since $C_{t}$
  hits $b$ before $a$ in this case, $T_{b}(y) < T_{a}(y)$ and
  $T_{a,b}(y) = T_{b}(y)$.}
\label{exit_hit_times}
\end{figure}

In order to quantitatively describe the dynamics of $C_{t}$, we
introduce two characteristic times, visually represented in
\figref{exit_hit_times}. We define the \emph{first hitting time
  $T_{u}(y)$ of $u$} to be the first time that $C_{t}$ with $C_{0} = y
\in [0,1]$ attains a given value $u$ in $[0,1]$, \emph{i.e.} $T_{u}(y)
= \inf_{t \geq 0} \{ C_{t} = u | \ C_{0} = y \}$. Similarly, we define
the \emph{first exit time $T_{a,b}(y)$ of $]a,b[$} to be the first
time that $C_{t}$ with $C_{0} = y \in [a,b]$ reaches either $a$ or
$b$, that is $T_{a,b}(y) = \inf_{t \geq 0} \{ C_{t} \notin \ ]a,b[ | \
C_{0} = y \}$. We note that $T_{a,b}(y) = \min\{T_{a}(y),T_{b}(y)\}$.

\subsection{Probability to hit $b$ before $a$}

To quantify the extent to which the emitter becomes excited as a
consequence of the measurements, it is convenient to introduce the
probability $P_{T_{b} < T_{a}}(y)$ that $C_{t}$, commencing from
$C_{0} = y \in \ [a,b]$, hits the value $b$ before
$a$. Following \cite{karlin1981}, we shall heuristically obtain a
differential equation for $P_{T_{b} < T_{a}}(y)$.

Let $P(y) \equiv P_{T_{b} < T_{a}}(y) $, and note that the probability
that $C_{t}$, starting from $C_{0} = y \in \ ]a,b[$ at time $t = 0$,
reaches $a$ or $b$ already after an infinitesimal time $dt$ is
vanishing. Thus, conditioned on the value of $C_{dt}$, the probability
to reach $b$ before $a$ at time $dt$ is $P(C_{dt})$ and we conclude
that the average value $\mean\bigl( P(C_{dt}) \bigr)$ over all
possible realizations of $C_{dt}$ starting at $C_{0} = y$ must equal
the initial probability $P(y)$. At the same time, we have from
\eqqref{SSE_C_het} or \eqqref{SSE_C_hom_opt} and the Ito rules that
$\mean(dC_{0}) = A(y)dt$ and $\mean(dC_{0}^{2}) = B(y)^{2}dt$.

Combining these two results and using that $dC_{0} = C_{dt} - C_{0} =
C_{dt} - y$, we obtain
\begin{align}
P(y) & = \mean\bigl( P(C_{dt}) \bigr) \nonumber \\
& = \mean\bigl( P( dC_{0} + y ) \bigr) \nonumber \\
& = \mean\Bigl( P(y) + P'(y) dC_{0} + \frac{1}{2} P''(y) dC_{0}^{2} \Bigr) \nonumber \\
& = P(y) + P'(y) A(y) dt + \frac{1}{2} P''(y) B(y)^{2} dt,
\label{diff_Py}
\end{align}
where we have Taylor expanded $P( dC_{0} + y )$ around $y$ to second
order in $dC_{0}$ and used the Ito rule $dW_{0}^{2} = dt$. Now,
subtracting $P(y)$ on both sides of \eqqref{diff_Py} and dividing
through by $dt$, we obtain the sought-after differential equation for
$P_{T_{b} < T_{a}}(y)$
\begin{align}
L_{y} P_{T_{b} < T_{a}}(y) = 0
\label{diff_P_Tb_<_Ta}
\end{align}
subject to the natural boundary conditions $P_{T_{b} < T_{a}}(a) = 0$
and $P_{T_{b} < T_{a}}(b) = 1$, where we have defined
\begin{align}
L_{y} = A(y)\frac{d}{dy} + \frac{1}{2} B(y)^{2}\frac{d^{2}}{dy^{2}},
\label{Ly}
\end{align}
For more details see Ch.~15.3 of \cite{karlin1981} and Thm.~4 in
Ch.~3.15 of \cite{gihman1972}.

The solution of \eqqref{diff_P_Tb_<_Ta} can be found explicitly by
direct integration to be
\begin{align}
P_{T_{b} < T_{a}}(y) = \frac{S(a,y)}{S(a,b)}
\label{P_Tb_<_Ta}
\end{align}
with $y \in [a,b]$, where
\begin{align}
S(u,v) = \int_{u}^{v} Q(z) dz,
& & Q(z) = e^{-2 \int^{z} \frac{A(z')}{B(z')^{2}} dz'}
\label{S_Q}
\end{align}
We note that for $Q$ and thus $S$ to be \emph{a priori} well-posed we
must require that the diffusion term $B(C_t)$ in
Eqs.~\eqref{SSE_C_het} and \eqref{SSE_C_hom_opt} obeys $B(C_t)^{2} >
0$, which is fulfilled for $0 < a < b < 1$.

\subsubsection{Heterodyne detection.}

For heterodyne detection, the integrals in \eqqref{S_Q} can be evaluated
analytically, and we obtain
\begin{align*}
Q(z) & = e^{-\frac{1}{z}} \frac{z}{1-z} \\
S(u,v) & = \Bigl[ \frac{1}{e} E_{1} \Bigl(\frac{1-z}{z}\Bigr) - e^{-\frac{1}{z}} z \Bigr]_{u}^{v}
\end{align*}
with limiting behavior
\begin{align}
S(0,b) < \infty, && S(a,1) = \infty && \text{for }  a,b \in [0,1[, \label{S_lim_het}
\end{align}
where $E_{1}(z) = \int_{z}^{\infty} \frac{e^{-z'}}{z'} dz'$ is the
$E_{1}$-function with $E_{1}(0) = \infty$ and $E_{1}(\infty) = 0$.

\subsubsection{Optimal homodyne detection.}

For homodyne detection, we obtain
\begin{align*}
Q(z) & = e^{- \frac{1}{2z} } \sqrt{\frac{z}{1-z}} \\
S(u,v) & = \Bigl[ - \sqrt{\frac{\pi}{2e}} \erf\Bigl( \sqrt{\frac{1-z}{2z}} \Bigr) - e^{-\frac{1}{2z}} \sqrt{z(1-z)} \Bigr]_{u}^{v}
\end{align*}
with
\begin{align}
S(0,b) < \infty, && S(a,1) < \infty && \text{for } a,b \in [0,1], \label{S_lim_hom}
\end{align}
where $\erf(z) = \frac{2}{\sqrt{\pi}}
\int_{0}^{z} e^{-z'^{2}} dz'$ is the error function with
$\erf(0) = 0$ and $\erf(\infty) = 1$.

\subsection{Attainability of ground and excited states and excitation probability}

While the unobserved, exponentially decaying emitter never exceeds its
initial excitation and only reaches the ground state asymptotically,
\emph{i.e.} after infinite time, during continuous photon count
detection, a fraction $C_{0} = y$ of the non-exponentially decreasing
trajectories jumps to the ground state at exponentially distributed,
finite times. For heterodyne or homodyne detection, however, it is not
\emph{a priori} clear whether the conditioned two-level system may
become fully excited or deexcited at finite times as a result of the
measurements.

While deferring a rigorous verification to \secref{char_times}, we
will now argue heuristically for the attainability of the extremal
values $C_t=0$ and $C_t=1$, and in the process obtain an analytical
expression for the excitation probability $P_{T_{u} < \infty}(y)$.

\subsubsection{Attainability of zero and unit excitation}

We suppose that the upper level population $C_{t} = \epsilon << 1$ at
time $t$. To attain the value $0$ in the next infinitesimal time step,
$C_{t + dt} = 0$, the measurement must induce a change in $C_{t}$
through a Wiener increment of size $dW_{t} = \frac{1 - \gamma dt}{k
  \sqrt{\gamma}\sqrt{\epsilon-\epsilon^{2}} }$ according to
Eqs.~\eqref{SSE_C_het} and \eqref{SSE_C_hom}, where $k =
\frac{2}{\sqrt{2}}$ for heterodyne detection and $k = 2$ for optimal
homodyne detection. In the limit $\epsilon \downarrow 0$, where the
excited state population becomes infinitesimally close to $0$, this
would require an arbitrarily large Wiener increment, $dW_{t}
\rightarrow \infty$. Consequently, the probability that the excited
state population of an atom with a non-zero initial excitation $y$
vanishes after a finite time is zero, and the first hitting time
$T_{0}(y)$ must be infinite with certainty,
\begin{align}
P_{T_{0} = \infty}(y) = 1 && \text{for } y \in \ ]0,1]
\label{P_T_0_=_inf}
\end{align}
For completeness, we note that a similar argument does not apply for
the attainability of the excited state, since the diffusive term
vanishes more slowly when $C_t$ approaches unity. If $C_{t} =
1-\epsilon$ is close to $1$, the Wiener increment required for
transitioning to $C_{t+dt} = 1$ after $dt$ is $dW_{t} \rightarrow -
\frac{1}{k\sqrt{\gamma}} ( \sqrt{\epsilon} + \gamma
\frac{dt}{\sqrt{\epsilon}} )$, which may occur with finite
probability in the limit $\epsilon \downarrow 0$.

Even though $C_{t}$ does not reach $0$ in finite time, the emitter
decays with certainty to the ground state in the long
time limit,
\begin{align}
\lim_{t \rightarrow \infty}C_{t} = 0
\label{lim_C}
\end{align}

\subsubsection{Excitation probability}

The above analysis yields an explicit formula for the probability that
$C_{t}$ attains any level population $u \in [0,1]$ in finite time,
\begin{align}
P_{T_{u} < \infty}(y) =
\begin{dcases}
\frac{S(0,y)}{S(0,u)} & \text{for } 0 < y \leq u \leq 1 \\
1 & \text{for } 0 < u \leq y \leq 1
\end{dcases},
\label{P_T_u_<_inf}
\end{align}
where the first case follows directly from \eqqref{P_Tb_<_Ta} by
putting $b = u$ and taking the limit $a \downarrow 0$ upon using
\eqqref{P_T_0_=_inf}. We note from Eqs.~\eqref{S_lim_het} and
\eqref{S_lim_hom} that this limit and the limit $u \uparrow 1$ are
well defined, and the formula also holds for $u = 1$. The second case
follows from the fact that $C_{t}$ tends continuously towards zero and
consequently passes all points in $]0,y]$ with unit probability.

\begin{figure}[htbp]
\centering
\includegraphics[scale = 1]{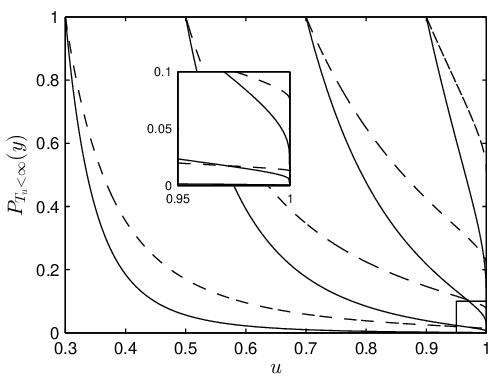}
\caption{\small The probability $P_{T_{u} < \infty}(y)$, given in
  \eqqref{P_T_u_<_inf}, that the upper level population $C_{t}$
  attains a specific value $u$ above its initial value $y = C_{0} =
  0.3,0.5,0.7,0.9$ (from left to right) for heterodyne (solid curves)
  and optimal homodyne (dashed curves) detection. The $y$-values are
  the same as in \figref{P_Tuy_num}. The smaller figure presents a
  magnified view of the region near unit excitation
  probability. $C_{t}$ only attains the value $1$ with non-vanishing
  probability for optimal homodyne detection while $C_{t}$ never
  reaches $1$ for heterodyne detection. }
\label{P_Tuy}
\end{figure}

For optimal homodyne detection, $S(0,1) < \infty$ according to
\eqqref{S_lim_hom}, and there is a \emph{non-vanishing} probability
that the upper level population $C_{t}$ of the emitter (initially
prepared in any superposition different from $\ket{g}$) attains
\emph{any} value $u$ in $]0,1]$ in \emph{finite} time. This is
illustrated in \figref{P_Tuy} for different initial level populations
$C_{0} = y$. In particular, the emitter has a finite probability to
become \emph{fully} excited as a result of the optimal, continuous
homodyne measurements.

For heterodyne detection, $S(0,1) = \infty$ from \eqqref{S_lim_het}
and, even though $C_{t}$ may attain any value $u \in \ ]0,1[$, the
excitation probability $P_{T_{u} < \infty}(y)$ drops to zero as $u
\uparrow 1$, and the emitter has vanishing probability to become fully
excited as seen in \figref{P_Tuy}. More specifically, since
$E_{1}(\frac{1-u}{u}) = -\ln(\frac{1-u}{u}) - \overline{\gamma} +
\mathcal{O}(1-u)$, where $\overline{\gamma} = 0.5772\ldots$ is the
Euler-Mascheroni constant, we see from \eqqref{P_T_u_<_inf} that
$P_{T_{u} < \infty}(y)$ varies as
$-\frac{S(0,y)}{\frac{1}{e}\ln(1-u)}$ in the limit $u \uparrow 1$. For
not too small $y$, examination of this asymptotic expansion (see
\figref{P_Tuy}) shows that $P_{T_{u} < \infty}(y)$ is finite for a
wide range of values of $u$, while dropping steeply to $0$ for $u$ very
close to unit excitation.

Finally, we note from \figref{P_Tuy} that compared to the optimal
homodyne detection case, an atom subject to heterodyne monitoring is
generally much less likely to become excited.

\section{Characteristic times for heterodyne and optimal homodyne detection}
\label{char_times}

\subsection{First exit time and rigorous justification of attainability}

We will now further characterize the dynamics of the observed emitter
by calculating the average of the first time $T_{a,b}(y)$ that the
conditioned excited state population $C_{t}$, commencing from $C_0= y
\in [a,b]$, hits either $a$ or $b$. As a by-product we obtain a rigorous
justification of the results on the attainability of zero and unit
excited state population in \secref{excitation_dynamics}.

In order to obtain a deterministic expression for $E(T_{a,b}(y)) =
\overline{T}_{a,b}(y)$, we abbreviate $T_{a,b}(y) = T(y)$ and observe
that the probability that the trajectory $C_{t}$, starting at time $t
= 0$ from $y \in \ ]a,b[$, will reach either $a$ or $b$ after an
infinitesimal time $dt$ is vanishing. Thus by the Markov property of
$C_{t}$, the first mean exit time of $]a,b[$ at time $dt$ is,
conditioned of the value of $C_{dt}$, given by $\overline{T}(C_{dt})$,
and it is clear that $\mean\bigl( \overline{T}(C_{dt}) \bigr) =
\overline{T}(y) - dt$, where the mean $\mean(\ldots)$ is taken over
all possible realizations of $C_{dt}$ commencing from $C_{0} = y$.

Using the Ito rules as in \eqqref{diff_Py}, we find
\begin{align*}
\overline{T}(y)
& = \mean\bigl( \overline{T}(C_{dt}) \bigr) + dt \\
& = \overline{T}(y) + \overline{T}'(y)A(y)dt + \frac{1}{2}\overline{T}''(y)B(y)^{2}dt + dt,
\end{align*}
and we conclude that $\overline{T}_{a,b}(y)$ solves the differential
equation
\begin{align}
L_{y} \overline{T}_{a,b}(y) = -1
\label{diff_Taby}
\end{align}
subject to the boundary conditions $\overline{T}_{a,b}(a) =
\overline{T}_{a,b}(b) = 0$, and with the differential operator $L_{y}$
given by the same expression \eqqref{Ly} as in the equation for
$P_{T_{b} < T_{a}}(y)$. See Ch.~15.3 of \cite{karlin1981} and
Ch.~5.2.7 of \cite{gardiner2004} for further details on the
derivation.

We can verify that the solution of \eqqref{diff_Taby} is
\begin{align*}
\overline{T}_{a,b}(y) = \frac{ S(y,b) R(a,y) - S(a,y) R(y,b) }{ S(a,b) }
\end{align*}
with $y \in [a,b]$, and $0 < a < b < 1$, where
\begin{align*}
R(u,v) = \int_{u}^{v} Q(z) K(z) dz, & & K(z) = \int^{z} \frac{- 2}{Q(z')B(z')^{2}}dz'
\end{align*}
and $S$ and $Q$ are specified in \eqqref{S_Q}.

\subsubsection{Heterodyne detection.}

For heterodyne detection, we obtain
\begin{align}
K(z) & = \frac{1}{\gamma} e^{\frac{1}{z}} \Bigl( 2 - \frac{2}{z} + \frac{1}{z^{2}} \Bigr) \label{K_het} \\
R(u,v) & = \frac{1}{\gamma} \Bigl[ \ln\Bigl(\frac{z}{1-z}\Bigr) - 2z \Bigr]_{u}^{v} \nonumber
\end{align}
with limiting behavior
\begin{align}
R(0,b) = \infty, && R(a,1) = \infty && \text{for } a,b \in \ [0,1] \label{R_lim_het}
\end{align}

\subsubsection{Optimal homodyne detection.}

For optimal homodyne measurements we obtain
\begin{align}
K(z) = {} & -\frac{1}{\gamma}\sqrt{\frac{1-z}{z^{3}}} (2z-1) e^{\frac{1}{2z}} + \frac{\sqrt{2\pi e}}{\gamma} \ \erfi \Bigl( \sqrt{ \frac{1-z}{2z} } \Bigl) \label{K_hom} \\
R(u,v) = {} & \frac{1}{\gamma} \bigl[  \ln(z) - 2z \bigr]_{u}^{v} \nonumber \\ & + \frac{\sqrt{2\pi e}}{\gamma} \int_{u}^{v} e^{-\frac{1}{2z}} \sqrt{\frac{z}{1-z}} \erfi \Bigl( \sqrt{ \frac{1-z}{2z} } \Bigr) dz \nonumber
\end{align}
where $\erfi(z) = \frac{2}{\sqrt{\pi}} \int_{0}^{z} e^{z'^{2}} dz'$ is
the imaginary error function with $\erfi(0) = 0$ and $\erfi(\infty) =
\infty$. We have not been able to evaluate the last integral in $R$ in
terms of standard functions, but we observe the limiting behavior as
the integrand in $R$ remains bounded on $[0,1]$
\begin{align}
R(0,b) = \infty, && R(a,1) < \infty && \text{for } a,b \in \ ]0,1], \label{R_lim_hom}
\end{align}

\subsubsection{Attainability of zero and unit excited state population}

We are now in the position to provide stringent verification of the
results in \secref{excitation_dynamics}. According to Thm.~1 in
Ch.~5.21 of \cite{gihman1972} and Lemma~6.2 in Ch.~15.6 of
\cite{karlin1981}, we have

\begin{thm}
\item If $P_{T_{b} < T_{0}}(y) > 0$ (i.e. if $S(0,b) < \infty$), then
  $\overline{T}_{0,b}(y) = \infty \Leftrightarrow R(0,y) = \infty
  \Leftrightarrow P_{T_{0} = \infty}(y) = 1$. \label{thm_1}
\end{thm}
\begin{thm}
\item If $P_{T_{b} < T_{0}}(y) > 0$ and $\overline{T}_{0,b}(y) =
  \infty$ (i.e. if $S(0,b) < \infty$ and $R(0,y) = \infty$), then
  either
\begin{inparaenum}[(i)]
\item $T_{0,b}(y) = \infty$ and $\lim_{t \rightarrow \infty} C_{t} =
  0$ or
\item $T_{0,b}(y) < \infty$ and $C_{T_{0,b}(y)} = b$.
\end{inparaenum} \label{thm_2}
\end{thm}

We verify from Eqs.~\eqref{S_lim_het}, \eqref{S_lim_hom},
\eqref{R_lim_het}, and \eqref{R_lim_hom} that the assumptions in
Thms.~\ref{thm_1} and \ref{thm_2} are fulfilled and \thmref{thm_1}
gives \eqqref{P_T_0_=_inf}.  \eqqref{lim_C} follows immediately if i)
of \thmref{thm_2} applies, while if ii) is true and the trajectory
$C_{t}$ hits $b$, since $C_{t}$ is Markovian, the theorem may be
applied again now with $C_{t}$ proceeding from $b$, \emph{i.e.} $y
\rightarrow b$ and $b \rightarrow b_{1} > b$. This argument may be
repeated, and as we see from the first case of \eqqref{P_T_u_<_inf},
the probability that $C_{t}$ starting from $b_{i}$ hits $b_{i+1} >
b_{i}$ is strictly smaller than $1$ and eventually i) of
\thmref{thm_2} will apply.

\subsection{Average excitation time}

Unlike the upper level population of the unobserved emitter, the
conditioned trajectory may spend a finite amount of time above its
initial excitation. More generally, $C_{t}$, commencing from $y \in
[a,1]$, may spend one or more periods of time in an interval $[\ell,r]
\ \subseteq [a,1]$ before exiting through $a$ for the first time. We
denote this time $T(y)$ and notice that its ensemble average
$\mean\bigl(T(y)\bigr) = \overline{T}(y)$ may be expressed as
\begin{align}
\overline{T}(y) = \mean \Bigl( \int_{0}^{T_{a}(y)} f(X_{t}) dt \Bigr)
\label{avTy_int}
\end{align}
where $f(z) = 1$ for $z \in [\ell,r]$ and $f(z) = 0$ otherwise. Here
$T_{a}(y)$ is the first time $C_{t}$ with $C_{0} = y$ hits the level
population $a$, as introduced in \secref{excitation_dynamics}. By
ultimately taking the limit $a \downarrow 0$ in \eqqref{avTy_int} and
employing that $T_{0}(y) = \infty$, we will obtain an expression for
the mean time spent by $C_{t}$ in any interval $[\ell,r] \subseteq
[0,1]$.

To obtain a differential equation for $\overline{T}(y)$, we break the
integral up into two contributions, $\overline{T}(y) = \mean (
\int_{0}^{dt} f(C_{t}) dt ) + \mean ( \int_{dt}^{T_{a}(y)} f(C_{t}) dt
)$. The first contribution is just $f(y)dt$. And observing that the
probability that $C_{t}$, commencing at time $t = 0$ from $y \in
]a,1]$, reaches $a$ during the infinitesimal time step $dt$ is
vanishing, we infer from the Markov property of $C_{t}$ that the
second contribution is equal to $\mean \bigl( \overline{T}(C_{dt})
\bigr)$, where the average is taken over all possible realizations of
$C_{dt}$ starting from $C_{0} = y$. Employing the Ito rules as in
\eqqref{diff_Py}, we find that
\begin{align*}
\overline{T}(y) & = f(y) dt + \mean \bigl( \overline{T}(C_{dt}) \bigr) \\
& = f(y) dt + \overline{T}(y) + \overline{T}'(y)A(y)dt + \frac{1}{2}\overline{T}''(y)B(y)^{2}dt,
\end{align*}
and $\overline{T}(y)$ solves
\begin{align}
L_{y}\overline{T}(y) = -f(y)
\label{diff_Ty}
\end{align}
subject to the boundary conditions $\overline{T}(a) = 0$ and
$\frac{d}{dy}\overline{T}(y)|_{y=1} = 0$, and where $L_{y}$ is given
in \eqqref{Ly}. Since $T_{a}(a) = 0$, the first boundary condition is
clear, while the second one follows by noticing that $C_{t}$ is
reflected at $C_{t} = 1$, and $T_{a}(y)$ and thus $\overline{T}(y)$
should have a maximum at $y = 1$. For more details, see Ch.~5.2.7 of
\cite{gardiner2004} for the special case $f(y) \equiv 1$, and Prob.~15
in Ch.~15 of \cite{karlin1981} for the general case.

With the prescribed boundary conditions, the solution of
\eqqref{diff_Ty} is
\begin{align}
\overline{T}(y) = \tilde{R}(a,y) - S(a,y) \tilde{K}(1),
\label{avTy}
\end{align}
where $\tilde{R}$ and $\tilde{K}$ are given by
\begin{align*}
\tilde{R}(u,v) = \int_{u}^{v} Q(z) \tilde{K}(z) dz, && \tilde{K}(z) = \int_{\ell}^{z} \frac{-2 f(z')}{Q(z')B(z')^2} dz',
\end{align*}
and $S$ and $Q$ are given in \eqqref{S_Q}. The explicit formulas are
lengthy, and we shall refrain from writing them here.

By taking the limit $a \downarrow 0$ in \eqqref{avTy} and remembering
that $T_{0}(y) = \infty$, we obtain an expression for the mean time
the trajectory $C_{t}$ resides in the interval $[\ell,r] \subseteq
[0,1]$. Since $C_{t}$ hits all level populations in $]0,1[$ with
non-vanishing probability and tends asymptotically towards $0$, this
time must be infinite if $\ell = 0$ and finite otherwise, as an
analysis of \eqqref{avTy} also reveals.

Putting $\ell = y$ and $r = 1$ in the limit $a \downarrow 0$,
$\overline{T}(y)$ yields the
average time spent by $C_{t}$ above its initial value $C_{0} = y$. We note that $\tilde{R}(0,y) = 0$ and $\tilde{K}(1) =
K(1) - K(y)$, where $K(1)$ is finite for both heterodyne and homodyne
detection according to Eqs.~\eqref{K_het} and \eqref{K_hom}, and we
find
\begin{align}
\overline{T}(y) = S(0,y)\bigl( K(y) - K(1) \bigr)
\label{avTex}
\end{align}
for $y \in [0,1]$ with $T(0) = 0$ (from the boundary condition) and
$T(1) = 0$, since, in case of heterodyne detection, $\overline{T}(y)$
goes like $-\frac{1}{\gamma} (1-y)^{2}\ln(1-y)$ when $y \uparrow 1$
according to the previously stated expansion of $E_{1}(\frac{1-y}{y})$.

\begin{figure}[htbp]
\centering
\includegraphics[scale = 1]{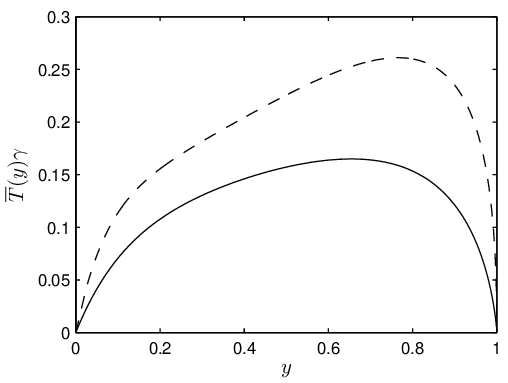}
\caption{The mean excitation time for heterodyne (solid curve) and
  optimal homodyne (dashed curve) detection given in \eqqref{avTex}.}
\label{Tex}
\end{figure}

In \figref{Tex}, we plot the mean excitation time for both heterodyne
and optimal homodyne detection, and observe that it is notably larger
for homodyne detection than for heterodyne detection for all initial
upper state populations $y \in ]0,1[$.

\subsection{Average time of decay}

In the Markov approximation assumed, monitoring does not modify the
average decay properties of the emitter, it merely unravels the master
equation solution into different quantum trajectories depending on the
detection scheme. Thus, on average $C_{t}$ equals the unconditioned
upper level occupation probability $\rho_{ee}(t)$ given in
\eqqref{rho_ee}. Starting from an initial excitation $\rho_{ee}(0) =
y$, the exponential decay of $\rho_{ee}(t)$ reaches an excited state
population $a$ at the time $t_{exp}(a)=\gamma^{-1}\ln \frac{y}{a}$.
Due to the fluctuations of individual trajectories, the first time
the conditioned trajectory $C_{t}$ hits a certain level $a$
below its initial value $C_{0} = y$ will be a fluctuating variable $T_{a}(y)$,
which may take both larger and smaller values than $t_{exp}(a)$, as
illustrated in \figref{C_t_pex}.

By taking $f(y) \equiv 1$ in \eqqref{diff_Ty}, we obtain a
differential equation for its average value $\mean \bigl( T_{a}(y)
\bigr) = \overline{T}_{a}(y)$
\begin{align}
L_{y}\overline{T}_{a}(y) = -1
\label{diff_T_ay}
\end{align}
with the boundary conditions $\overline{T}_{a}(a) = 0$ and
$\frac{d}{dy}\overline{T}_{a}(y)|_{y=1} = 0$.

The solution of \eqqref{diff_T_ay} is
\begin{align}
\overline{T}_{a}(y) & = R(a,y) - S(a,y)K(1)
\label{avT_ay}
\end{align}
for $0 < a \leq y \leq 1$. In the limit $a \downarrow 0$ we get from
Eqs.~\eqref{R_lim_het} and \eqref{R_lim_hom} that $\overline{T}_{0}(y)
= \infty$, as expected since the first hitting time of $0$ is infinite
by \eqqref{P_T_0_=_inf}. For heterodyne detection, the expansion
of $E_{1}(\frac{1-y}{y})$ implies that both $R(a,y)$ and $S(a,y)K(1)$
diverge like $- \frac{1}{\gamma}\ln(1-y)$ as $y \uparrow 1$, thus
leaving $\overline{T}_{a}(1)$ finite.

\begin{figure}[htbp]
\centering
\includegraphics[scale = 1]{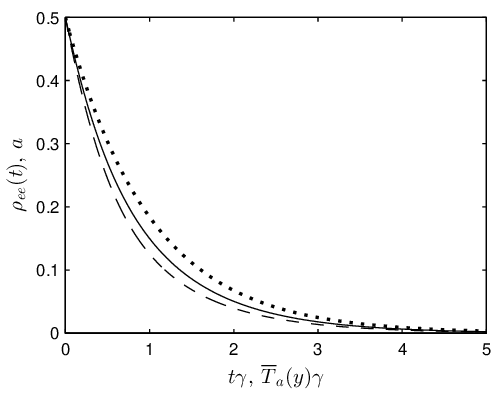}
\caption{Comparison of exponential decay (dotted curve), $\bigl(
  t\gamma,\rho_{ee}(t) \bigr)$, with the average time to reach excited
  state probabilities $a$ below $y$, $\bigl(
  \gamma\overline{T}_{a}(y),a \bigr)$, during heterodyne (solid curve)
  and optimal homodyne (dashed curve) detection. The initial
  excitation is $\rho_{ee}(0) = y = 0.5$. The two-level system
  conditioned on continuous, heterodyne or optimal homodyne
  measurements reach lower excitation faster on average than the
  unconditioned master equation solution.}
\label{T_ay}
\end{figure}

In \figref{T_ay}, we compare the average exponential decay in
\eqqref{rho_ee} with the average first passage of excitation levels
below the initial population for both heterodyne and optimal homodyne
detection given in \eqqref{avT_ay}.  The figure shows that on average
$C_{t}$ conditioned on heterodyne detection reaches lower values
faster than $\rho_{ee}(t)$. It is interesting to observe that homodyne
detection, which leads to excursions to higher degrees of excitation,
also shows the faster average approach to low excitation levels as
compared to heterodyne measurements.

\section{Conclusion}
\label{conclusion}

The current interest in investigating the properties and achievements
of conditioned dynamics are stimulated by the prospects of combining
measurement back-action and feedback for control of quantum
systems. We have investigated the stochastic evolution of a two-level
emitter subject to continuous homodyne and heterodyne detection. In
particular, we have found that the emitter may reach any excited level
population, including unit excitation, as a result of in-phase,
optimal homodyne measurements. Homodyne detection $\pi/2$ out of phase
with the emitting dipole, on the other hand, cannot increase the
excitation beyond its initial value, while homodyne detection with
other local oscillator phases as well as heterodyne detection may lead
to any excitation probability strictly below unity.

These results do not violate energy conservation since the initial
state is assumed not to be an energy eigenstate. The results obtained
are characteristics of the measurement process and of its ability to
distinguish and unravel the dynamics into different conditioned
trajectories. A projective measurement in the energy eigenstate basis
would, indeed, yield the fully excited state with probability equal to
the initial excited state population. It is interesting, however, that
heterodyne and homodyne measurements are based on the detection of the
radiation emitted by the source, and that, in order to provide this
field, the system has to loose energy. In a photon counting
experiment, the detection of an emitted quantum of energy would be
associated with a jump of the emitter into its ground state, while
intervals of no counts would cause a continuous reduction of the
excited state population \cite{dalibard1992}, and the excitation would
not increase above its initial level.

Conditioned dynamics has been demonstrated to be a useful and
efficient way to prepare quantum states which are difficult or even impossible to reach
by unitary time evolution, and our
analysis of how even a very simple system can be prepared in an
excited state by monitoring its decay, points to rich opportunities
for further exploration of the achievements of different detection
schemes.

\bibliography{bibliography}

\end{document}